\documentclass{article}
\usepackage[utf8]{inputenc} 

\usepackage{geometry} 
\usepackage{graphicx} 
\usepackage{amsmath}
\usepackage{amssymb}
\usepackage{booktabs} 
\usepackage{array} 
\usepackage{paralist} 
\usepackage{verbatim} 
\usepackage{subcaption} 
\usepackage{float}
\usepackage{fancyhdr} 
\usepackage{physics}

\usepackage{tikz}
\usetikzlibrary{matrix,chains,positioning,decorations.pathreplacing,arrows}
\usepackage{qcircuit}
\usepackage{float}

\usepackage{sectsty}
\usepackage{ulem}
\allsectionsfont{\sffamily\mdseries\upshape} 
\usepackage[nottoc,notlof,notlot]{tocbibind} 
\usepackage[titles,subfigure]{tocloft} 

\newtheorem{remark}{Remark}

\usepackage{xcolor}
\newcommand{\red}{\color{red}}

\title{Continuous Variable Quantum Perceptron}

\author{F. Benatti$^{1,2}$, S. Mancini$^{3,4}$, S. Mangini$^1$\\
\small ${}^1$Dipartimento di Fisica, Universit\`a di Trieste, Strada Costiera 11, I-34151, Trieste, Italy\\
\small ${}^2$Istituto Nazionale di Fisica Nucleare, Sezione di Trieste, Strada Costiera 11, I-34151, Trieste, Italy\\
\small ${}^3$School of Science and Technology, University of Camerino
I-62032 Camerino, Italy\\
\small ${}^4$Istituto Nazionale di Fisica Nucleare, Sezione di Perugia, Via A. Pascoli, I-06123 Perugia, Italy
}

\date{\null}

\begin{document}
\maketitle

\begin{abstract}

We present a model of Continuous Variable Quantum Perceptron (CVQP) whose architecture implements a classical perceptron. 
The necessary non-linearity is obtained via measuring the output qubit and using the measurement outcome as input to an activation function.
The latter is chosen to be the so-called ReLu activation function by virtue of its practical feasibility and the advantages it provides in learning tasks. The encoding of classical data into  realistic finitely squeezed states and the use of superposed (entangled) input states for specific binary problems are discussed. 

\end{abstract}

\section{Introduction}

Quantum Machine Learning brings together Machine Learning and Artificial Intelligence on one side and Quantum Information and Computation on the other one.
Both sides  have recently witnessed a series of breakthroughs which herald them as fundamental ingredients of future technologies. As a consequence a steadily growing amount of research has been focussing upon whether these two fields could benefit from each other. Many generalizations of quantum architectures for machine learning tasks and, vice versa, classical machine learning aided quantum computational architectures are currently being explored and tested \cite{Briegel, Wossing, Wittek, Schuld, Loyd}.

In the following we focus on the study of a possible quantum implementation of a classical perceptron, the backbone of any learning algorithm, in a Continuous Variable (CV) quantum architecture \cite{QuantumCV}, based  on harmonic oscillator-like degrees of freedom, instead of \textit{discrete}, \hbox{spin-like} variables. Some generalizations of perceptron models in the quantum regime have already been proposed \cite{Macchiavello, Cao}, but most of them are built on top of a discrete quantum system, made of \textit{qubits}. Since a vast class of learning algorithms requires calculation of derivatives, which in turn needs continuous quantities to be evaluated, it is important to explore a possible quantum implementation of a perceptron 
that be continuous in nature. 

Models of quantum perceptrons based on the pseudo eigen-projectors of position operators are easily mathematically constructible; however, for all practical purposes one has to investigate 
how to concretely  implement these formal  architectures by means of approximations based on square-integrable states with fairly well, but not perfect continuous localization properties.
In doing this, one is then forced to consider the cost in energy that need be spent to counterbalance non-perfect state localization and the inaccuracies that it introduces.
We shall investigate the relation between the energy cost and the probability of classification error in the case of the AND and XOR rules.
In addition, it is worth underlining that the perceptron activation function considered in this work is the ReLu (Rectified Linear Unit) which recently proved to be an optimal choice for learning taks \cite{ReLu, DeepLearning}, and we shall introduce a measurement protocol devoted to its implementation. 
We shall also show that, in tackling the XOR and AND problems, no apparent advantage results from linear superpositions and entanglement 
possibly present in the input states of the quantum perceptron.

The structure of the work is as follows. In Section \ref{sec:Classical Perceptron} all the necessary concepts about classical perceptrons are reviewd, in Section \ref{sec:Continuous Quantum Perceptron} the proposed model for a continuous quantum perceptron with the ReLu activation function is presented. In Section \ref{sec:Gaussian Input States} realistic and feasible input states for the quantum perceptron are considered and in Section \ref{sec:AND} 
the performance for the AND problem is studied while in Section \ref{sec:XOR} the XOR problem is addressed using linear superpositions. 
Finally, in Section \ref{sec:Conclusions}, an outlook of possible further research directions is drawn.

\section{Classical Perceptron}
\label{sec:Classical Perceptron}

\begin{figure}[ht]
\centering
\begin{tikzpicture}[
init/.style={
  draw,
  circle,
  inner sep=2pt,
  font=\Huge,
  join = by -latex
},
squa/.style={
  draw,
  inner sep=2pt,
  font=\Large,
  join = by -latex
},
start chain=2,node distance=13mm
]
\node[on chain=2] 
  (x2) {$x_2$};
\node[on chain=2,join=by o-latex] 
  {$w_2$};
\node[on chain=2,init] (sigma) 
  {$\displaystyle\Sigma$};
\node[on chain=2,squa,label=above:{\parbox{2cm}{\centering Activation \\ function}}]   
  {$f$};
\node[on chain=2,label=above:Output,join=by -latex] 
  {$y$};
\begin{scope}[start chain=1]
\node[on chain=1] at (0,1.5cm) 
  (x1) {$x_1$};
\node[on chain=1,join=by o-latex] 
  (w1) {$w_1$};
\end{scope}
\begin{scope}[start chain=3]
\node[on chain=3] at (0,-0.75cm) 
  (xp) {\vdots};
\node[on chain=3] 
  (wp) {\vdots};
\end{scope}
\begin{scope}[start chain=3]
\node[on chain=3] at (0,-1.5cm) 
  (xn) {$x_n$};
\node[on chain=3,label=below:Weights,join=by o-latex] 
  (wn) {$w_n$};
\end{scope}
\node[label=above:\parbox{2cm}{\centering Bias \\ $b$}] at (sigma|-w1) (b) {};

\draw[-latex] (w1) -- (sigma);
\draw[-latex] (wn) -- (sigma);
\draw[o-latex] (b) -- (sigma);

\draw[decorate,decoration={brace,mirror}] (x1.north west) -- node[left=10pt] {Inputs} (xn.south west);
\end{tikzpicture}
\caption
{Schematic representation of the mathematical model of a perceptron.}
\label{fig:McCullochNeuron}
\end{figure}
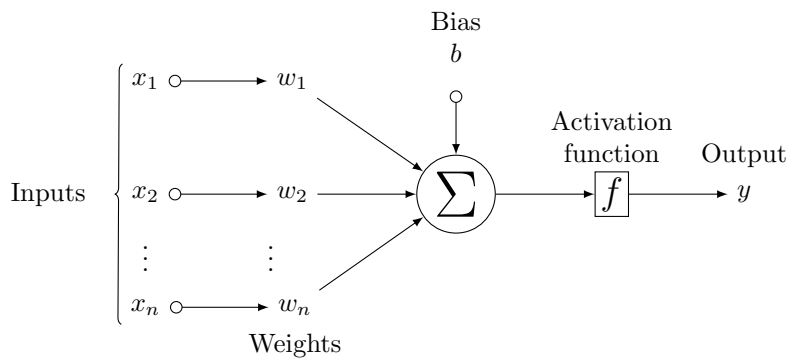

By perceptron it is meant a mathematical model that mimics the functioning of a natural perceptron. Given an input $\vec{x} = (x_1, \hdots, x_n)$ with $x_i \in \mathbb{R}$, a perceptron computes an affine transformation with real parameters $\vec{w}=(w_1, \hdots, w_n)$ and $b$, called \textit{weights} and \textit{bias}, respectively:
\begin{equation}
\label{affunct}
\vec{x}\longmapsto z:=\vec{x}\cdot \vec{w} + b\  .
\end{equation}
Subsequently, the perceptron evaluates on the output  $z$ an \textit{activation function} $f: \mathbb{R} \rightarrow \mathbb{R}$, eventually yielding 
the final result $y:=f(z)$.

There exist different possible activation functions, some being more computationally efficient and others more biologically inspired as the hyperbolic tangent $f(z)= ({\rm e}^{z}-{\rm e}^{-z})/({\rm e}^{z}+{\rm z}^{-x})$, or the sigmoid function $f(z)=1/(1+{\rm e}^{-z})$. More recently however, the nonlinear function known as  Rectified Linear Unit, ReLu for short, has proved to be a very good candidate in learning procedures~\cite{ReLu, DeepLearning}:
\begin{equation}
\text{ReLu}(z):=z^+=\max(0,z)\qquad\forall z\in\mathbb{R}\ .
\label{eq:ReLu}
\end{equation}

In order to illustrate how a classical perceptron can be used, consider the AND binary function on pairs $(x_1,x_2)\in\{-1,1\}^2$ whose truth table is reported in the following 
Table~\ref{tab:ANDtruth}.
 \begin{table}[ht]
\centering
\begin{tabular}{c|c|c}
$x_1$ & $x_2$ & $x_1$ AND $x_2$ \\ \hline
-1 & -1 & 0 \\
1 & -1 & 0 \\
-1 & 1 & 0 \\
1 & 1 & 1 \\
\end{tabular}
\caption[Truth table for the AND function]{Truth table of the AND problem.}
\label{tab:ANDtruth}
\end{table}
The AND function maps the four pairs $(x_1,x_2)$ into $1$ if and only if $x_1=x_2=1$ otherwise the output is $0$.
One says that such a function is computed by the perceptron if the pairs $(x_1,x_2)\neq (1,1)$ are univocally associated with the output $0$ and the pair $(1,1)$ with the output $1$.
Choosing the weights $w_1=w_2=1$ and the bias $b=-1$ one gets
$$
z=x_1\,+\,x_2\,-\,1=\left\{\begin{matrix}
+1&\hbox{if}\quad(x_1,x_2)&=&(1,1)\cr
-1&\hbox{if}\quad(x_1,x_2)&=&(1,-1)\cr
-1&\hbox{if}\quad(x_1,x_2)&=&(-1,1)\cr
-3&\hbox{if}\quad(x_1,x_2)&=&(-1,-1)\cr
\end{matrix}\right.\ ,
$$
whence $\hbox{ReLu}(z)=1$ only in the first case, whence the pair $(1,1)$ is then univocally separated  from the other three ones for which $\hbox{ReLu}(z)=0$.

Unfortunately, the hope for a full classification of the whole of $16$  binary logical functions by a classical perceptron is hindered by the exclusive-OR (XOR) logical function.
Its truth table is reported in Table~\ref{XORTab}. Indeed, the XOR function outputs $1$ only if only one of the input  values  is $1$. Such a classification problem
cannot be solved by a one-layer classical perceptron as it is not linearly separable; namely, the points $(1,-1)$, $(-1,1)$ in the $(x_1,x_2)$ plane cannot be univocally 
separated (by means of a line) from the points $(1,1)$, $(-1,-1)$.

\begin{table}[ht]
\centering
\begin{tabular}{c|c|c}
$x_1$ & $x_2$ & $x_1$ XOR $x_2$ \\ \hline
-1 & -1 & 0 \\
1 & -1 & 1 \\
-1 & 1 & 1 \\
1 & 1 & 0 \\
\end{tabular}\ .
\caption[Modified truth table for the XOR function]{Truth table for the XOR function.}
\label{XORTab}
\end{table}


\section{Continuous Variable Quantum Perceptron Model}
\label{sec:Continuous Quantum Perceptron}

The continuous quantum degrees of freedom we are going to use in the following are of photonic type described by annihilation and creation operators $\hat a_i$, $\hat a^\dag_i$, $[\hat a_i\,,\hat a_i^\dagger]=1$, or more conveniently  by the quadrature position and momentum operators 
\begin{equation}
\label{quadratures}
\hat x_i=\frac{\hat a_i\,+\,\hat a_i^\dag}{\sqrt{2}}\ ,\quad \hat p_i=\frac{\hat a_i\,-\,\hat a_i^\dag}{i\sqrt{2}}\ ,\qquad[\hat x_i\,,\,\hat p_j]=i\,\delta_{ij}\ .
\end{equation}
Within the formalism of continuous quantum computation  by means of photons, 
the pseudo-eigenstates of $\hat{x}_i$, $\hat x_i\vert x_i\rangle=x_i\,\vert x_i\rangle$, 
play the same role as does the computational basis $\{\ket{0}, \ket{1}\}$ in discrete qubit  systems; namely,
any action on generic optical states can be described in terms of them. 

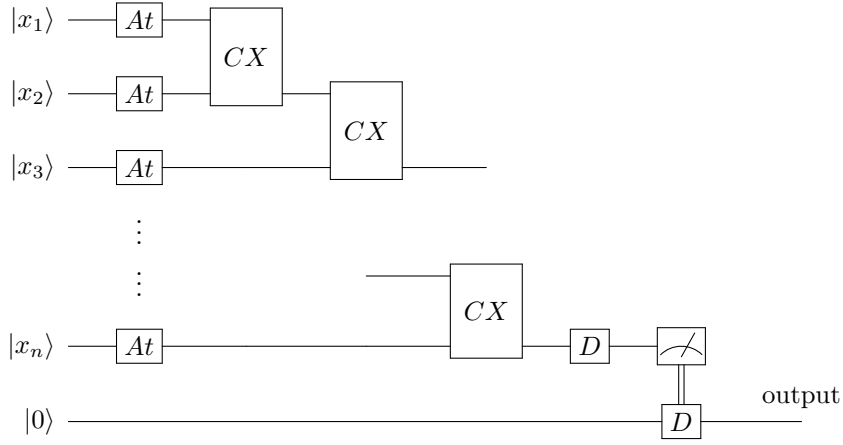
\begin{figure}[ht]
\[ \Qcircuit @C=1.8em @R=1.5em {
      \lstick{\ket{x_1}} & \gate{At} & \multigate{1}{CX} \\
      \lstick{\ket{x_2}} & \gate{At} & \ghost{CX}& \multigate{1}{CX}\\
      \lstick{\ket{x_3}} & \gate{At} & \qw & \ghost{CX} & \qw \\
       & \vdots &  \\
       & \vdots & & & \multigate{1}{CX} \\
      \lstick{\ket{x_n}} & \gate{At} & \qw & \qw & \ghost{CX} & \gate{D} & \meter \cwx[1]\\
      \lstick{\ket{0}} & \qw & \qw & \qw & \qw & \qw & \gate{D} &  \qw &\ustick{\text{output}} \qw \\
      } \]
      \caption{Scheme for a continuous valued quantum perceptron}
      \label{fig:QN}
      \end{figure}

A quantum circuit implementing the behaviour of a classical perceptron when acting on a quantum continuous optical input  is summarized by the scheme in Fig.~\ref{fig:QN}. 

A multimode input state consisting of a common eigenstate $\ket{x_1, x_2, \hdots, x_n}$ of the quadrature operators $\hat{x}_j$, $j=1, 2, \hdots, n$,
$\hat{x}_j\ket{x_1, x_2, \hdots, x_n}=x_j\ket{x_1, x_2, \hdots, x_n}$, undergoes three successive steps: in the first one each component of the input state  is affected by a series of attenuators \textit{At} which multiply each eigen-positions $x_j$ by a real scaling factor $\eta_j$ with $|\eta_j|\leq 1$. During the second step the attenuated eigen-positions are recursively added  by means of suitably Controlled Addition $CX$ gates; finally,  in the last step, by means of a Displacement gate $D$, a bias is added to the last eigen-position.  After these three steps, the activation function is implemented by acting on the last position eigen-state by performing a threshold ideal homodyne measurement. 

The initial register consists of the $n$ input state encoding the data to be processed and of an additional ancillary system, initialized in the pseudo-position eigenstate $\ket{0}$.
Its role is  to correctly implement the activation function and to propagate the result of the quantum perceptron. The visual representation of the later as a quantum circuit is shown in Figure \ref{fig:QN} and 
can be schematically summarized as follows:

\begin{itemize}

\item \textbf{Attenuation}\\
The attenuation process, 
$ \Qcircuit @C=1.2em @R=1em { & \gate{At} & \qw } $,
implements the multiplication of each eigen-position $x_j$ by  a corresponding weight $|\eta_j|\leq 1$:
\begin{equation}
    \ket{x_1,x_2, \hdots,x_n}\rightarrow \ket{\eta_1 x_1,\eta_2 x_2, \hdots,\eta_n x_n}\ .
    \label{eq::At}
\end{equation}
Such a transformation can be obtained by means of squeezing unitary gates. 
In fact, the single qumode squeezing operator 
\begin{equation}
\label{Sq1}
S(r)=\exp(i\,r\,(\hat x\hat p+\hat p\hat x))=\exp(r\,(a^2-(a^\dag)^2))\ ,\quad r \in \mathbb{R}^+\ ,
\end{equation} 
yields 
\begin{eqnarray}
\label{Sq2a}
S^\dag(r)\,\hat x\,S(r)&=&{\rm e}^{-2r}\,\hat x\ ,\qquad S^\dag(r)\,\hat p\,S(r)={\rm e}^{2r}\,\hat p\\
\label{Sq2b}
S^\dag(r)\,\hat a\,S(r)&=&\hat a\, \cosh(2r)\,-\,\hat a^\dag\,\sinh(2r)\ .
\end{eqnarray}
From the first expression in~\eqref{Sq2a} it follows that
\begin{equation}
\label{Sq3}
S(r)\ket{x}={\rm e}^{-r}\ket{{\rm e}^{-2r}x}\ .
\end{equation}
Setting $\eta={\rm e}^{-2r}$, the following transformation is obtained:
\begin{equation}
\ket{x_1,x_2, \hdots,x_N}\rightarrow \sqrt{\eta_1 \eta_2 \cdots \eta_n} \ket{\eta_1 x_1,\eta_2 x_2, \hdots,\eta_n x_n}\ .
\label{eq:Attenuation}
\end{equation}
Thus, the strengths $\eta_j$ of the attenuation processes  implement the weights of the classical perceptron. Notice that the attenuation process performed by the squeezing operator $S(r)$ in~\eqref{Sq1} yields $0\leq \eta={\rm e}^{-2r} \leq 1$. In order to implement negative weights, one can use a simple Phase Shift gate $R(\phi) :=\exp[i\phi \hat{a}^\dagger \hat{a}]$ with phase $\phi=\pi$. In this case the Attenuation gate would comprise both a Squeezing and a Rotation gate, $\Qcircuit @C=1em @R=1em { & \gate{At} & \qw }\ =\  \Qcircuit @C=1em @R=1em { & \gate{S} & \gate{R} & \qw }$, leading to the compound transformation:
\begin{equation}
\ket{x} \rightarrow \sqrt{\eta}\,\ket{\eta x} \rightarrow \sqrt{\eta}\,\ket{-\eta x}\ .
\end{equation}
\item \textbf{Controlled Addition}\\
The controlled addition gate $\Qcircuit @C=1em @R=1em {  
	 & \multigate{1}{CX} \\
     & \ghost{CX}& \qw}$ processes two inputs by keeping the first one unaltered and adding  it to the second one.
The operator responsible for a such a process can be easily seen to be the following one: 
\begin{equation}
\label{eq:CX}
CX:= \exp[-\frac{i}{\hbar}\hat{x}_1 \otimes \hat{p}_2]\ ,\qquad CX(\ket{x_1}\otimes\ket{x_2})=\ket{x_1}\otimes\ket{x_1+x_2}\ .
\end{equation}
There are many possible implementations of such a gate, via two suitable beam-splittings that come before and after a squeezing gate~\cite{StrawberryFields} or via quantum-non demolition 
processes~\cite{QND1, QND2, QND3, QND4}.     
The combined action on the attenuated state of the $n-1$ $CX$ gates of the circuit in Fig.~\ref{fig:QN}  is then given by
\begin{equation}
\begin{split}
    \ket{\eta_1 x_1,\eta_2 x_2, \hdots,\eta_n x_n} & \rightarrow \ket{\eta_1 x_1,\eta_1 x_1+\eta_2 x_2, \hdots,\eta_n x_n}  \rightarrow \hdots \\ 
    & \hdots \rightarrow \ket{\eta_1 x_1,\eta_1 x_1+\eta_2 x_2, \hdots,\sum_{i=1}^{n} \eta_i x_i}\ .
\end{split}
\end{equation}
Namely, they iteratively sum the position  of one qumode system to the following one, eventually obtaining the sought after weighted sum as eigen-position of the last qumode.

\item  \textbf{Insertion of a bias}\\
Using the $n$-th qumode as a reading mode, a bias $b\in\mathbb{R}$ can be added to its position by means of a Displacement operator 
$\Qcircuit @C=1em @R=1.5em { & \gate{D} & \qw }$:
\begin{eqnarray}
\label{Displ1a}
D(b)&=&\exp(-i\,\,b\,\hat p)=\exp(b\frac{\hat a^\dag-\hat a}{\sqrt{2}})\\
\label{Displ1b}
D^\dag (b)\,\hat x\,D(b)&=&\hat x\,+\,b\ ,\quad D^\dag(b)\,\hat a\,D(b)=\hat a\,+\frac{b}{\sqrt{2}}\ .
\end{eqnarray}
Then, from the first equality in~\eqref{Displ1b}, one derives that the initial position eigenstate $\vert x_n\rangle$ of the last qumode is finally transformed into 
\begin{equation}
\label{AffineTrans}
D(b)\vert\sum_{i=1}^{n} \eta_i x_i\rangle=\vert  \sum_{i=1}^{n} \eta_i x_i+b\rangle
\end{equation}
exactly as demanded by the affine transformation~\eqref{affunct}.

\item \textbf{Activation function}\\
Having encoded  the affine transformation into the last qumode eigen-position, the last step consists in implementing the ReLu activation function \eqref{eq:ReLu} via a threshold measurement on it, $\Qcircuit @C=1em @R=1.5em { & \meter & \qw }$: it mimics the non-linear behaviour of the classical perceptron. Such a task can be performed by an ideal homodyne measurement with POVM elements given by the pseudo-projector operators $P_y=\ket{y}\bra{y}$ onto the pseudo-position eigenstates $\ket{y}$.
Once performed on the last qumode when its state $\hat \rho$, such a measurement yields $y$ with probability $P(y)=\mel{y}{\hat{\rho}}{y}$. 
Upon receiving $y$ as measurement outcome, using the Displacement gate~\eqref{Displ1a} one displaces the  pseudo-position eigenstate $\ket{0}$ of the ancilla in the following way:
\begin{equation}
\label{laststep}
     \ket{0} \rightarrow 
  \begin{cases} 
   \ket{0} & \text{if } y \leq 0 \\
   \vert y\rangle & \text{if } y > 0
  \end{cases}
\end{equation}
Such a conditional action, $\Qcircuit @C=1em @R=1.5em {  
	   & \meter \cwx[1]\\
       & \gate{D} & \qw}$ implements the ReLu activation function \eqref{eq:ReLu} and eventually encodes the final result into state of the ancilla qumode. 
       In the following, we shall consider more realistic scenarios where position pseudo-eigenstates are substituted by fairly well localized normalizable states. Consequently, the ideal pseudo-eigenstate $\ket{y}$ in~\eqref{laststep} will also be substituted by the displaced vacuum state $D(y)\ket{0}$, where now $\ket{0}$ denotes the vacuum state, such that
       $\hat{a}^\dagger\hat{a}\ket{0}=0$.

\end{itemize}

\begin{remark}
\label{rem1}
Other possibilities for the activation function could be implemented substituting the threshold measurement protocol with suitable nonlinear and non gaussian gates, such as the Kerr gate. Unfortunately, due to the non interacting nature of photons and to the lack of sufficiently strong nonlinear materials with low absorption~\cite{EffSimQuantum}, such a gate is very difficult to implement physically. For this reason, measurement induced nonlinearities provide feasible and valuable alternatives. In addition, by means of the described conditional measurement protocol, it is possible to implement the ReLu activation function, which recently provided promising advantages for deep learning tasks.
Notice that with this model the quantum implementation of a classical perceptron  requires at least $n$ qumodes for the signals to add up and one more ancillary qumode to implement the activation function. In addition, more modes may be necessary for the implementation of the gates used in the computation.
\end{remark}

\section{Gaussian Input States}
\label{sec:Gaussian Input States}

Continuous position eigenstates can not provide actual physical input states to the quantum perceptron because they correspond to Dirac deltas over the continuum of eigen-positions, and are thus not accessible as such  in laboratory. However, they can be approximated by means of square integrable Gaussian states fairly well localized around the  eigen-positions. 
Hence it is important to investigate the action of the quantum circuit~\eqref{fig:QN}  on Gaussian wave packets of the form
\begin{equation}
\ket{\psi_j} = \frac{1}{(\pi \Delta^2_j)^{1/4}} \int d q_j \  {\rm e}^{-\frac{(q_j - x_j)^2}{2 \Delta^2_j}}\ket{q_j}\ ,
\label{eq:WavePacket}
\end{equation}
which are Gaussian weighted normalized superpositions of  pseudo-eigenstates $\ket{q_j}$, of the position quadrature $\hat x_j$, where the classical datum $x_j$ is encoded as center of the Gaussian weights with width $\Delta_j$. 
Such states are obtained by acting on the vacuum state $\ket{0}$ first with a Squeezing gate as in~\eqref{Sq1} and then with Displacement gate
as in~\eqref{Displ1a}.
Indeed,  in position representation, $\langle q\vert 0\rangle=(\exp(-q^2/2)/\sqrt[4]{\pi}$,
so that the resulting Displaced-Squeezed vacuum state 
is \cite{Dahl}:
\begin{equation}
\langle q\vert\,D(b)S(r)\vert 0\rangle=\frac{1}{{(\pi {\rm e}^{-2r})} ^{1/4}}\, \exp\left(-\frac{(q-b)^2}{2 {\rm e}^{-2r}}\right)\ ,
\label{eq:gaussianDSV}
\end{equation}
which reduces to~\eqref{eq:WavePacket} when ${\rm e}^{-r}=\Delta_j$ and $b=x_j$. Notice that, when $r$ becomes large, these states approximate a Dirac delta 
around $x_j$. In the following we shall denote by $\ket{x_j,\Delta_j}$ the Displaced-Squeezed vacuum states $D(x_j)S(-\log\Delta_j)\ket{0}$.

It then follows that the input state to the quantum circuit in Fig.~\ref{fig:QN} is
\begin{equation}
\ket{\Psi}=\bigotimes_{j=1}^n\ket{\psi_j}=\prod_{j=1}^n \frac{1}{(\pi \Delta_j^2)^{1/4}}\int dq_j\,{\rm e}^{-\frac{(q_j - x_j)^2}{2 \Delta^2_j}}
\bigotimes_{j=1}^n\ket{q_j}\ .
\label{eq:Nwavepacket}
\end{equation}

After applying  the Attenuation and CX gates of the quantum circuit in FIg.~\ref{fig:QN}, the outcome probability provided by the ideal homodyne detection reads 
(see Appendix \ref{App:GaussianCalculation})
\begin{equation}
P (y,\vec{x})=\frac{1}{\sqrt{\pi \sum_{j=1}^N \eta^2_j\Delta^2_j }} \ {\rm e}^{-\frac{(y-b-\sum_{j=1}^N \eta_j x_j)^2}{\sum_{j=1}^N \eta^2_j\Delta^2_j}}\ ,
\qquad  \vec{x}=(x_1,x_2,\ldots, x_N)\ .
\label{eq:FinalPy}
\end{equation}
It thus corresponds to a normalized Gaussian, centered around  the result $\vec{w}\cdot\vec{x}+b$ of the affine transformation in~\eqref{affunct}. 
Depending on the actual outcome of the measurement process, the ancilla qumode will then be displaced by $y=f(z)$ and the final output read out.

However, differently from the case of the unnormalizable position-eigenstates, 
after homodyne measurements, states with not sharp eigen-positions  
yield all possible $y\in\mathbb{R}$ with a given probability distribution that in turn determines a probability of error associated to
a wrong pattern classification by the quantum perceptron.
Indeed, suppose that $a>0$ is the correct answer for a given input state $\ket{\Psi_a}$ with associated outcome probability density $P(y)$. Actually, the ReLu activation function 
is able to discriminate between two kinds of states only, namely either the vacuum state $\ket{0}$, or Gaussian coherent states $D(a)\ket{0}$ with $a>0$ (see the discussion just before Remark~\ref{rem1}). Then,
\begin{equation}
\label{errprob}
P_{err}(\vec{x}):=\int_{-\infty}^{0}P(y,\vec{x})\ dy\ ,
\end{equation}
represents the worst probability of miscalculation, that is the probability of obtaining a negative value from the homodyne measurement which in turn leads to an output ancilla 
qumode in the vacuum state, thus misclassifying the input state.

Furthermore, obtaining Displaced-Squeezed states has an energy cost that can be computed as the mean value of the number operator $\hat{n}=\hat{a}^\dagger\hat{a}$ with respect to 
such a state which is proportional to the electromagnetic energy content of an optical mode.
Hence, as a figure of merit, the \textit{energy cost} of obtaining a Displaced-Squeezed vacuum state from the vacuum state 
can be taken to be the difference of the mean values of the number operator computed with respect to those states, namely:
\begin{equation}
\label{energy}
E(x_j, \Delta_j):= \mel{x_j, \Delta_j}{\hat{n}}{x_j, \Delta_j} -\mel{0}{\hat{n}}{0}= \frac{|x_j|^2}{2}\,+\,\frac{(1-\Delta_j^2)^2}{4\Delta_j^2}\ .
\end{equation}
Such an an energy cost diverges when going to large squeezing parameters $r$, namely when the Squeezed Displaced vacuum approximates a Dirac delta around the displacement 
parameter.
The argument of above then shows that a bound on the expendible energy unavoidably degrades  the performances of the quantum perceptron  
introducing a non-vanishing probability of error. It is thus important to study the trade-off between the energy spent for a better spatial localization and the corresponding  lowering of the error probability.


\section{Quantum computation of the AND function}
\label{sec:AND}

In order to study  the performances of the quantum circuit in Fig.~\ref{fig:QN} as a model of quantum perceptron, we now probe it against the AND function introduced in Section~\ref{sec:Classical Perceptron} and relate the classification errors to the energy bound used in the encoding of the classical data into the input states. 

\begin{figure}[ht]
\[ \Qcircuit @C=1.8em @R=1.5em {
      \lstick{\ket{x_1;\Delta_1}} & \gate{At} & \multigate{1}{CX} \\
      \lstick{\ket{x_2;\Delta_2}}& \gate{At} & \ghost{CX} &  \gate{D} & \meter \cwx[1] \\
      \lstick{\ket{0}} & \qw & \qw & \qw & \gate{D} & \qw &  \ustick{x_1\ \text{AND}\ x_2} \qw \\
      } \]
      \caption[Scheme for implementing and AND function using DSV states]{Scheme for implementing and AND function using Displaced-Squeezed vacuum states.}
      \label{fig::QNand2}
      \end{figure}
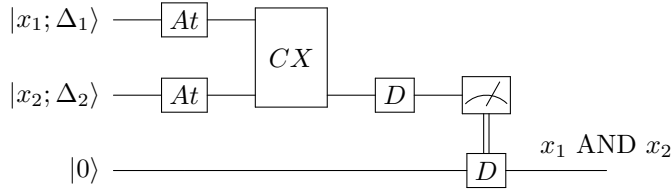

In Fig.~\ref{fig::QNand2} it is depicted a quantum circuit that might be used to compute the AND function: in order to correctly classify the input pairs $(x_1,x_2)$, it must ouput the vacuum state $\ket{0}$ if one or both the inputs $x_i$ are negative or the displaced vacuum 
$\ket{y}$, with $y>0$, if both are positive. 

Could one work with position pseudo-eigenstates,  then choosing attenuators $\eta_1=\eta_2=1$ and bias $b=-1$ exactly as the classical weights and bias would correctly 
implements the AND function as with a classical perceptron.
Instead, in an actual experimentally feasible context, the use of Displaced-Squeezed input states \eqref{eq:gaussianDSV} and the corresponding probability distribution 
$P(y,x_1,x_2)$ in~\eqref{eq:FinalPy}, yields the results reported in Table \ref{tab:ANDresults}, where the probability of error  $P_{err}(x_1,x_2)$ is defined as in~\eqref{errprob} namely as the integral of the probability  $P(y,x_1,x_2)$ over all possible results leading to misclassification.

\begin{table}[ht]
\centering
\begin{tabular}{ccccc}
\toprule
\multicolumn{2}{c}{Input}  & & \multicolumn{2}{c}{ $P_{err}(x_1,x_2)$}  \\ 
$x_1$  & $x_2$ & & $r=0$ & $r=1$  \\ \cline{1-2}\cline{4-5} \\[-1em]
    -1           &  -1             &  &  $0.13\%$       &     $\sim 10^{-14}\%$  \\
    -1           &  +1            &  &  $15.8\%$       &     $0.3\%$  \\
    +1           &  -1            &  &  $15.8\%$        &     $0.3\%$  \\ 
    +1           &  +1            &  &  $15.8\%$       &     $0.3\%$  \\ \bottomrule
\end{tabular}
\caption[Performance of the quantum perceptron]{Probability of misclassification, for each possible input, depending on the squeezing parameter 
$r=-\log\Delta$, where $\Delta$ is the width of the input Displaced-Squeezed vacuum states.}
\label{tab:ANDresults}
\end{table}

Evidently,  a correct implementation of the AND function requires that the input states be squeezed in order to reduce the probability of errors. Already with squeezing factor $r=1$, a good implementation is obtained. With this choice of squeezing parameter, the worst case scenario has a probability of error of just $0.3\%$. Thus, this quantum neuron could be actually used for a safe enough implementation of a single AND function. However, the evaluation of more complicated functions, by means of a network composed of multiple copies of such a quantum neuron, could be hindered by the accumulation of single neuron errors, though this effect could be controlled using an higher squeezing.
The energy cost due to encoding the classical inputs into Displaced-Squeezed states, to which there contribute the mean values of the number operators $\hat{n}=\hat{a}^\dagger\hat{a}$,  amounts to \cite{KokLovett}:
\begin{equation}
E_{tot}=E(x_1,\Delta_1)+E(x_2,\Delta_2)= \frac{|x_1|^2+|x_2|^2}{2}\,+\,\frac{(1-\Delta_1^2)^2}{4\Delta_1^2}\,+\,\frac{(1-\Delta_2^2)^2}{4\Delta_2^2}\ .
\label{eq:EnergyAND}
\end{equation}

\begin{remark}
\label{rem2}
Squeezing with $\Delta\ll 1$ is indeed very energy consuming, thus keeping $\Delta\simeq 1$ is preferable for an efficient classification. As a comparison, by encoding the classical input pair into Coherent states (without squeezing them) with displacements $|x_1|=|x_2|=2$, and using a lower bias, $b=-2$, then, the greatest probability of error amounts to $2.27\%$, which is about one order of magnitude larger than by encoding through Displaced-Squeezed vacuum states, for as much the same energy cost. In fact, while for the Displaced-Squeezed encoding ($|x_1|=|x_2|=1$ and $r=1\Leftrightarrow\Delta=1/e$) the energy cost amounts to \eqref{eq:EnergyAND} $E_{tot}\sim 3.76$, using Displaced states with squeezing parameter $r=0\Leftrightarrow\Delta=1$ and displacements
$|x_1|=|x_2|=2$ leads to an energy $E_{tot}=4$.
\end{remark}

\begin{figure}[H]
\centering 
\includegraphics[width=\linewidth]{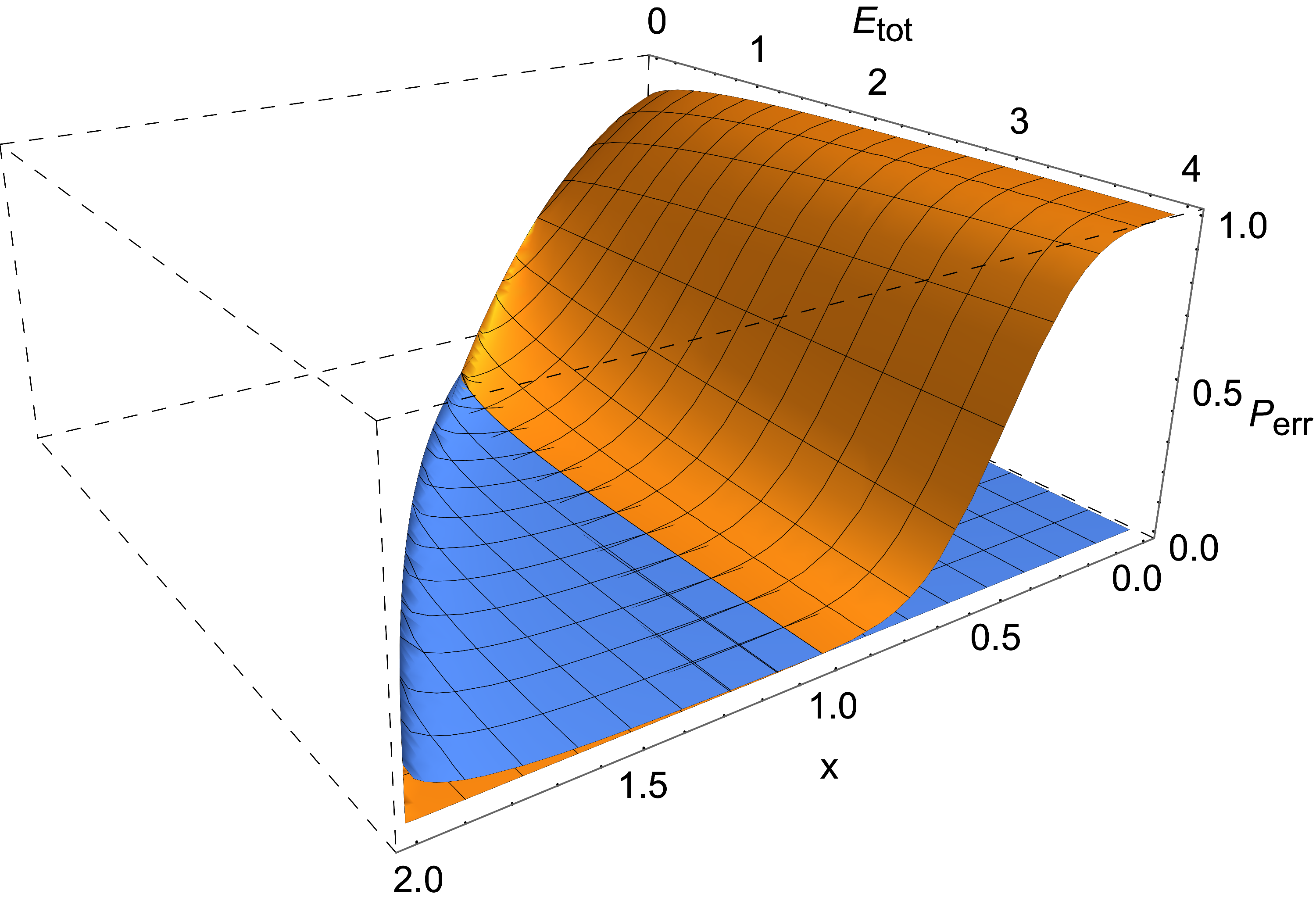}
\caption{$P_{err}^+(x,E_{tot})$ (yellow surface) and $P_{err}^-(x,E_{tot})$ (blue surface)
 as functions of the displacement $x$ and energy cost of the state $E_{tot}$}.
\label{fig:Perr_En_AND}
\end{figure}

Further insights about the energy cost are shown in the above Fig.~\ref{fig:Perr_En_AND}, where we assumed $|x_1|=|x_2|=|x|$ and allowed for the encoding of the inputs $(\pm1,\pm1)$ and $(\pm1,\mp1)$ by continuous variables $(x,x)$ and $(x,-x)$ with $\Delta_1=\Delta_2=\Delta$. Then, one can invert \eqref{eq:EnergyAND} to have
\begin{equation}\label{eq:Delta}
\Delta^2=\left(1+E_{tot}-|x|^2\right){\red -}\sqrt{\left(1+E_{tot}-|x|^2\right)^2-1} \,
\end{equation}
whence, setting $\eta_1=\eta_2=1$ and $b=-1$,  we can express \eqref{eq:FinalPy} as
\begin{align}
P^+(y,x,E_{tot})&=\frac{1}{\sqrt{2\,\pi\,\Delta^2}}\exp\left[-\frac{(y+1-2x)^2}{2\Delta^2}\right], 
\qquad x_1=x_2=x \\
P^-(y,x,E_{tot})&=\frac{1}{\sqrt{2\,\pi\, \Delta^2}}\exp\left[-\frac{(y+1)^2}{2\Delta^2}\right], 
\qquad\qquad x_1=-x_2=x 
\end{align}
Leaving outside the case corresponding to the inputs $(-1,-1)$ which according to Table \ref{tab:ANDresults} shows low probability of error, according to \eqref{errprob} we will then have two probabilities of error, $P_{err}^+(x,E_{tot})$, corresponding to the case $(+1,+1)$,
and $P_{err}^-(x,E_{tot})$, corresponding to the case with inputs with different signs.
Figure \ref{fig:Perr_En_AND} shows that until the energy is below one, it would be better to use the energy entirely  for displacing the vacuum by $|x|$ with no energy consumption due to squeezing,
$\Delta=1$. 
On the other hand, once the displacement reaches $|x|=1$, then the eventual extra energy can be used to squeeze the displaced vacuum so to achieve lower probability of error represented by $P_{err}^+(x,E_{tot})$ . 
\medskip

Choosing different values for the parameters involved in the computation (including attenuators and bias) allows for different possible encodings, but in any case a compromise between accuracy and energy is needed. 
In all cases, the instance of the AND function  indicates that the quantum circuit in Fig.~\ref{fig::QNand2} provides a realistic quantum implementation of a classical perceptron. 


\section{Quantum computation of the XOR function}
\label{sec:XOR}

As we have seen in Section~\ref{sec:Classical Perceptron}, the exclusive-OR (XOR) logical function represents a limit for a classical perceptron. 
It is then worthwhile investigating whether the quantum circuit implementation of a classical perceptron can do better in such a case by exploiting quantum state superpositions and/or
non-classical correlations.
As we shall see, quantum superpositions lead to 
a probabilistic implementation  of the XOR function. 
We start with analyzing position pseudo-eigenstates in order to get an idea of suitable inputs.
Using the quantum circuit in Figure~\ref{fig::QNand2}, with attenuators set to $\eta_1=1, \eta_2=-1$, bias set to $b=-1$, and classical inputs $(x_1,x_2)$ 
encoded into the superposition
\begin{equation}
\label{eq:psiinput}
\ket{\psi_{in}}=\ket{x_1,x_2}+\ket{x_2,x_1}\ ,
\end{equation}
we get
$$
\ket{\psi_{in}}\rightarrow \ket{\psi_{out}}=\ket{x_1, x_1-x_2-1}+\ket{x_2, x_2-x_1-1}\ .
$$
For input values $x_1=x_2=x$, $\ket{\psi_{out}}=\ket{x, -1}+\ket{x, -1}=2\ket{x, -1}$, whence the only possible result  from the homodyne measurement on the second qumode 
is correctly negative. On the contrary, when $x_1=-x_2=x$, then $\ket{\psi_{out}}=\ket{x,2x-1}+\ket{-x,-2x-1}$.  In this case, for both $x_1=1,\ x_2=-1$ and $x_1=-1,\ x_2=1$, the second output mode is left in an equally weighted mixture of states $\ket{-3}$ and $\ket{1}$.
Then the homodyne measurement statistics will result
half of the times in the negative domain and half in the positive one, thus leading to a probabilistic and imperfect classification. Therefore, when an input with $x_1=x_2$ is passed to the perceptron, it never misclassifies it, while when an input $x_1=-x_2$ is presented, it has a $50\%$ probability of being misclassified. 

\begin{figure}[H]
  \centering
  \includegraphics[width=\linewidth]{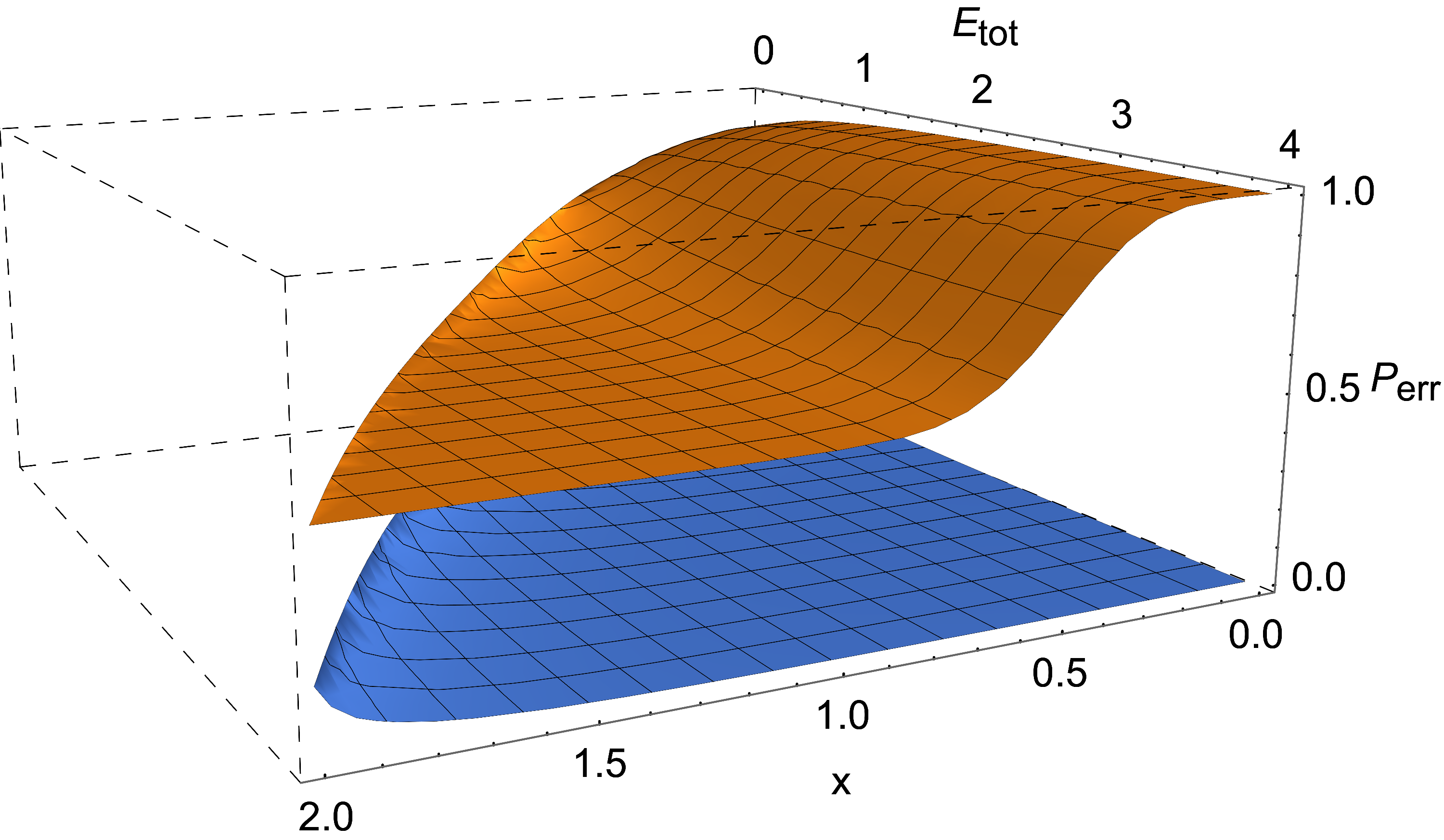}
  \caption{ $P_{err}^+(x,E_{tot})$ (yellow surface) and $P_{err}^-(x,E_{tot})$ (blue surface)
 as functions of the displacement $|x|$, and energy cost of the state $E_{tot}$.}
   \label{fig:XORdifferent}
\end{figure}

Let us make this more rigorous by resorting to normalizable states.
Taking for simplicity equal widht for the gaussian states, that is $\Delta_1=\Delta_2$, a feasible realization of the superposition state \eqref{eq:psiinput} reads:
\begin{eqnarray}
\label{superpos1}
\ket{\psi_{in}}&=&{\frac{1}{C}\Big(\ket{x_1,\Delta}\otimes\ket{x_2,\Delta}\,+\,\ket{x_2,\Delta}\otimes\ket{x_1,\Delta}\Big)}\\
\label{superpos2}
&=&\frac{1}{(C \pi \Delta^2)^{1/2}}\int d q_1 d q_2\left( {\rm e}^{-\frac{(q_1-x_1)^2}{2\Delta^2}}{\rm e}^{-\frac{(q_2-x_2)^2}{2\Delta^2}}+{\rm e}^{-\frac{(q_1-x_2)^2}{2\Delta^2}} {\rm e}^{-\frac{(q_2-x_1)^2}{2\Delta^2}}   \right)\ket{q_1, q_2}\ ,
\end{eqnarray}
where $C$ is a normalization constant that amounts to
\begin{equation}
C=2\left( 1+ {\rm e}^{-\frac{(x_1-x_2)^2}{2\Delta^2}}\right)\ .
\end{equation}
A discussion on the realizability of somehow similar states (just squeezed states are replaced by coherent ones) can be found in \cite{Gilchrist_2004}.
Applying the circuit with attenuators $\eta_1=1$ and $\eta_2=-1$, the probability $P(y)$ of the ideal measurement becomes (see Appendix \ref{app:XOR}):
\begin{equation}
\begin{split}
P(y) =  \frac{1}{C \sqrt{2 \pi\Delta^2} }\left( {\rm e}^{-\frac{(y-b-x_1+x_2)^2}{2\Delta^2}} + {\rm e}^{-\frac{(y-b-x_2+x_1)^2}{2\Delta^2}} 
 + \  2{\rm e}^{-\frac{(x_1-x_2)^2}{2\Delta^2}}{\rm e}^{-\frac{(y-b)^2}{2\Delta^2}}\right).
\end{split}
\label{eq:PyXOR}
\end{equation}
Furthermore, setting $b=-1$ we can express \eqref{eq:PyXOR} as
\begin{align}
P^-(y,x,E_{tot})&=\frac{1}{\sqrt{2\pi  \Delta^2}} {\rm e}^{-\frac{(y+1)^2}{2\Delta^2}}, 
\hspace{7.2cm} x_1=x_2=x \\
P^+(y,x,E_{tot})&=\frac{{\rm e}^{-\frac{(y+1-2x)^2}{2\Delta^2}}
+{\rm e}^{-\frac{(y+1+2x)^2}{2\Delta^2}}
+2{\rm e}^{-\frac{(y+1)^2+4x^2}{2\Delta^2}}}{(1+{\rm e}^{-2x^2/\Delta^2})\sqrt{8\pi \Delta^2}}\ ,
\hspace{3cm} x_1=-x_2=x 
\end{align}
with $\Delta$ as in \eqref{eq:Delta}. As much as in Section~\ref{sec:AND}, we will then have two probabilities of error $P_{err}^+(x,E_{tot})$ and $P_{err}^-(x,E_{tot})$.
They are depicted in Figure \ref{fig:XORdifferent}. 
We observe that $P^+(y,x,E_{tot})$  is always greater than $1/2$, which means that the quantum perceptron fails the classification most of the time when $x_1=-x_2$. 
In the limit of infinite energy, $r\rightarrow \infty$, and with $|x|=1$, one correctly obtains $P_{err}^+\rightarrow 1/2$, that is the same behaviour explained before with pseudo-position eigenstates. 

Assuming the 4 possible inputs come with equal probability $1/4$, with increasing energy costs, the quantum circuit tends to answer correctly 75\% of the times, which is the same as for a classical perceptron which can always correctly classify at least 3 of the 4 possible inputs. 
The main difference between the two perceptrons is that while the classical version always misclassifies at least one input, the quantum perceptron 
acts probabilistically on the inputs, misclassifying some of them at times. 

\section{Conclusions}
\label{sec:Conclusions}

A continuous-variable quantum circuit model has been proposed that shows the capability of correctly implementing the desired classification power achievable by a classical perceptron. The use of realistic input states has been considered and shown to provide an efficient trade-off  between a proper classification and the energy cost for those inputs.
 
The quantum nature of the circuit implementing the classical perceptron makes it possible the use of linear superpositions of quantum states; these latter, that have no classical counterpart, could in principle lead to classification advantages. However, our results indicate no quantum improvement when dealing with binary problems like the XOR and AND functions. Furthermore, for the AND function, there are preliminary evidences that also the use of an entangled encoding consisting of a two mode squeezed vacuum  does not bring advantages both in terms of probability and of energy cost, with respect to using non-entangled two single mode squeezed states. 

Actually, this should not come as a surprise, because the model proposed has been conceived as a direct copy of the classical model of a perceptron, and no further assumptions or ingredients have been used to achieve some kind of quantum advantage. In addition, the proposed quantum circuit falls within the hypothesis of the extended Gottesman Knill theorem \cite{EffSimQuantum}, which makes the quantum circuit acting on Gaussian inputs efficiently simulatable on a classical computer. Interestingly, no advantages come even when going beyond Gaussian inputs through linear superpositions of Displaced-Squeezed vacuum states. 

Therefore, it appears that quantum advantages are not easily achieved and the particular problems addressed in the manuscript are just an evidence of this fact. However, the possibility of 
processing more inputs at the same time through linear superpositions might require more sophisticated implementations of the perceptron non-linearity to make good use of them.
Nonetheless, it is important to underline that the proposed model works correctly both with position eigenstates and also with the realistic case of Gaussian states. It does then prove 
to be a good candidate for a quantum perceptron as a backbone of neural networks and machine learning techniques based on continuous quantum computation. 



\appendix

\section{Measurement statistics}
\label{App:GaussianCalculation}

Given the total input state in~\eqref{eq:Nwavepacket}, 
using the relation~\eqref{eq:Attenuation}, each of its constituent Gaussian wavepacket~\eqref{eq:WavePacket} is attenuated into 
\begin{equation}
At(\eta_j) \ket{\psi_j}= \frac{1}{(\pi \Delta^2_j)^{1/4}} \int d q_j \  {\rm e}^{-\frac{(q_j - x_j)^2}{2 \Delta^2_j}} At(\eta_j)\ket{q_j}\\
 = \frac{\sqrt{\eta_j}}{(\pi \Delta^2_j)^{1/4}} \int d q_j \  {\rm e}^{-\frac{(q_j - x_j)^2}{2 \Delta^2_j}}\ket{\eta_j q_j}\ .
\end{equation}
As a consequence, upon acting with all  attenuators and controlled addition gates on $\vert\Psi\rangle$ one obtains (for simplicity the case with bias $b=0$ is considered):
\begin{equation}
\begin{split}
    \ket{\tilde{\Psi}}=\prod_{j=1}^n \left( \frac{\sqrt{\eta_j}}{(\pi \Delta^2_j)^{1/4}}\right) \int dq_1 dq_2 & \hdots dq_n \  {\rm e}^{-\frac{(q_1 - x_1)^2}{2 \Delta^2_1}}{\rm e}^{-\frac{(q_2 - x_2)^2}{2 \Delta^2_2}}\cdots \ {\rm e}^{-\frac{(q_n - x_n)^2}{2 \Delta^2_N}}\\
    & \times \ket{\eta_1 q_1,\eta_1 q_1 +\eta_2 q_2, \hdots, \sum_{i=1}^n \eta_i q_i}\ .
\end{split}
\label{eq:PsiTransformed}
\end{equation}
Introducing the new variables:
\begin{equation}
\tilde{q}_j=\sum_{k=1}^j\eta_k q_k\ , \qquad j=1,2,\ldots, n\ , \quad q_{i+1}=\frac{\tilde{q}_{i+1}-\tilde{q}_i}{\eta_{i+1}}\ ,
\label{eq:VariableChange}
\end{equation}
with Jacobian $J= \prod_{j=1}^{n}\eta^{-1}_j$, the state in \eqref{eq:PsiTransformed}, can be recast as:
\begin{equation}
\begin{split}
    \ket{\tilde{\Psi}}=\frac{1}{\prod_{j=1}^n(\pi \Delta^2_j \eta^2_j)^{1/4}} \int d\tilde{q}_1 d\tilde{q}_2 & \hdots d\tilde{q}_n \  {\rm e}^{-\frac{(\tilde{q}_1 - \eta_1 x_1)^2}{2 \eta^2_1 \Delta^2_1}}{\rm e}^{-\frac{(\tilde{q}_2 -\tilde{q}_1 -\eta_2 x_2)^2}{2 \eta^2_2\Delta^2_2}} \cdots \times \\
    & \times {\rm e}^{-\frac{(\tilde{q}_n-\tilde{q}_{n-1} - \eta_n x_n)^2}{2 \eta^2_n \Delta^2_n}}\ket{ \tilde{q}_1, \tilde{q}_2, \hdots, \tilde{q}_n}\ .
\end{split}
\label{eq:PsiTransformed2}
\end{equation}
Such a  state is correctly normalized as required by the unitary action of the attenuators and $CX$ gates, the corresponding orthogonal projector 
$P_{\tilde{\Psi}}=\vert\tilde{\Psi}\rangle\langle\tilde{\Psi}\vert$ being
\begin{equation}
\begin{split}
    P_{\tilde{\Psi}}=& \frac{1}{\prod_{j=1}^N(\pi \Delta^2_j \eta^2_j)^{1/2}}
    \int dq_1 \hdots dq_n \ dq'_1 \hdots dq'_n {\rm e}^{-\frac{(q_1 - \eta_1 x_1)^2}{2 \eta^2_1 \Delta^2_1}}{\rm e}^{-\frac{(q_2 -q_1 -\eta_2 x_2)^2}{2 \eta^2_2\Delta^2_2}}\cdots \times\\
    &\times {\rm e}^{-\frac{(q_n-q_{n-1} - \eta_n x_n)^2}{2 \eta^2_n \Delta^2_n}} \, {\rm e}^{-\frac{(q'_1 - \eta_1 x_1)^2}{2 \eta^2_1 \Delta^2_1}}{\rm e}^{-\frac{(q'_2 -q'_1 -\eta_2 x_2)^2}{2 \eta^2_2\Delta^2_2}} \cdots \ {\rm e}^{-\frac{(q'_n-q'_{n-1} - \eta_n x_n)^2}{2 \eta^2_n \Delta^2_n}} \times \\
    &\times\ket{q_1,q_2, \hdots, q_n}\bra{q'_1,q'_2, \hdots, q'_n}\ .
\end{split}
\end{equation}
Since the affine transformation is encoded  into the $n$-th qumode, the remaining ones can be traced out thus yielding the $n$-th qumode reduced density matrix
\begin{equation}
\begin{split}
    \hat{\rho}_n & = \text{Tr}_{1, \hdots, \ n-1}( P_{\tilde{\Psi}})=\int dq''_1 \hdots dq''_{n-1} \mel{q''_1, \hdots, q''_{n-1}}{\rho}{ q''_1, \hdots, q''_{n-1}} \\
    		   & = \frac{1}{\prod_{j=1}^{n} (\pi \Delta^2_j \eta^2_j)^{1/2}} \int d q_1 d q_2 \hdots d q_{n-1} 							 \ {\rm e}^{-\frac{(q_1-\eta_1 x_1)^2}{\eta^2_1 \Delta^2_1}}\cdots \ {\rm e}^{-\frac{(q_{n-1}-q_{n-2}-\eta_{n-1} x_{n-1})^2}										{\eta^2_{n-1} \Delta^2_{n-1}}} \times \\
              &\hskip 3cm
        			\times \int d q_n d q'_n {\rm e}^{-\frac{(q_n-q_{n-1}-\eta_{n} x_{n})^2}{2 \eta^2_{n}											\Delta^2_{n}}} {\rm e}^{-\frac{(q'_n-q_{n-1}-\eta_{n} x_{n})^2}{2 \eta^2_{n}
      			    \Delta^2_{n}}}\ket{q_n}\bra{q'_n}\ .    
\end{split}
\label{eq:rhoN}
\end{equation}
One can check that $\text{Tr}(\hat{\rho}_n)=1$. Then, the remaining step to perform  is  the ideal  homodyne measurement on the transformed $n$-th mode. Using the orthogonality relation between position eigenstates $\braket{q}{y}=\delta(q-y)$, the probability $P(y)$ to obtain a given outcome $y \in \mathbb{R}$ is given by:
\begin{equation}
\begin{split}
    P(y) & =\mel{y}{\hat\rho_n}{y} \\
         	& = \frac{1}{\prod_{j=1}^{n} (\pi \Delta^2_j \eta^2_j)^{1/2}} \int d q_1 d q_2 \hdots d q_{n-1} \ 							{\rm e}^{-\frac{(q_1-\eta_1 x_1)^2}{\eta^2_1\Delta^2_1}} \cdots \ {\rm e}^{-\frac{(q_{n-1}-q_{n-2}-\eta_{n-1} x_{n-1})^2}										{\eta^2_{n-1}\Delta^2_{N-1}}} \times \\
         	& \hskip 4.5cm  \times {\rm e}^{-\frac{(y-q_{n-1}-\eta_n x_n)^2}{\eta^2_n\Delta^2_n}}\ .
\end{split}
\label{eq:Py}
\end{equation}

By  first integrating with respect to $dq_{n-1}$ and then applying iteratively the relation
\begin{equation}
    \int_{-\infty}^{\infty} dx \ {\rm e}^{-\frac{(y-x-a)^2}{b}} \ {\rm e}^{-\frac{(x-z-c)^2}{d}} =\sqrt{\frac{{\pi}}{\frac{1}{b}+\frac{1}{d}}}{\rm e}^{-\frac{(y-a-c-z)^2}{b+d}}\ ,
    \label{eq:Gauss}
\end{equation}
one gets
\begin{equation}
P (y)=\frac{1}{\sqrt{\pi \sum_{j=1}^N \eta^2_j\Delta^2_j }} \ {\rm e}^{-\frac{(y-\sum_{j=1}^N \eta_j x_j)^2}{\sum_{j=1}^N \eta^2_j\Delta^2_j}}\ .
\end{equation}
Finally, inserting the bias, the measurement outcome probability distribution reads
\begin{equation}
P (y)=\frac{1}{\sqrt{\pi \sum_{j=1}^N \eta^2_j\Delta^2_j }} \ {\rm e}^{-\frac{(y-b-\sum_{j=1}^N \eta_j x_j)^2}{\sum_{j=1}^N \eta^2_j\Delta^2_j}}\ .
\end{equation}

\section{Measurement statistics for the XOR binary function}
\label{app:XOR}

The state
\begin{equation}
\ket{\psi_{in}}=\frac{1}{(C \pi \Delta_1 \Delta_2)^{1/2}}\int d q_1 d q_2\left( {\rm e}^{-\frac{(q_1-x_1)^2}{2\Delta_1^2}}{\rm e}^{-\frac{(q_2-x_2)^2}{2\Delta_2^2}}+{\rm e}^{-\frac{(q_1-x_2)^2}{2\Delta_1^2}} {\rm e}^{-\frac{(q_2-x_1)^2}{2\Delta_2^2}}   \right)\ket{q_1, q_2}\ ,
\end{equation}
by application of the attenuators ($\eta_1=1,\ \eta_2=-1$) and controlled addition, is transformed into
\begin{equation}
\frac{-1}{(C \pi \Delta_1 \Delta_2)^{1/2}}\int d q_1 d q_2\left( {\rm e}^{-\frac{(q_1-x_1)^2}{2\Delta_1^2}}{\rm e}^{-\frac{(q_2-x_2)^2}{2\Delta_2^2}}+{\rm e}^{-\frac{(q_1-x_2)^2}{2\Delta_1^2}} {\rm e}^{-\frac{(q_2-x_1)^2}{2\Delta_2^2}}   \right)\ket{q_1,q_1- q_2}\ ,
\end{equation}
and changing integration variables $q_1 \rightarrow q_1$, $q_1-q_2 \rightarrow q_2$, it is eventually obtained
\begin{equation}
\ket{\tilde{\psi}_{in}}=\frac{1}{(C \pi \Delta_1 \Delta_2)^{1/2}}\int d q_1 d q_2\left( {\rm e}^{-\frac{(q_1-x_1)^2}{2\Delta_1^2}}{\rm e}^{-\frac{(q_2-q_1+x_2)^2}{2\Delta_2^2}}+{\rm e}^{-\frac{(q_1-x_2)^2}{2\Delta_1^2}} {\rm e}^{-\frac{(q_2-q_1+x_1)^2}{2\Delta_2^2}}   \right)\ket{q_1, q_2}\ .
\end{equation}
Then, tracing out the first mode yields 
\begin{equation}
\begin{split}
\Tr_1\left(\ket{\tilde{\psi}_{in}}\bra{\tilde{\psi}_{in}}\right) = &\frac{1}{C \pi \Delta_1 \Delta_2}\int d q_1 d q_2 d q'_2 \left( {\rm e}^{-\frac{(q_1-x_1)^2}{2\Delta_1^2}}{\rm e}^{-\frac{(q_2-q_1+x_2)^2}{2\Delta_2^2}}+{\rm e}^{-\frac{(q_1-x_2)^2}{2\Delta_1^2}} {\rm e}^{-\frac{(q_2-q_1+x_1)^2}{2\Delta_2^2}} \right) \\
& \times \left( {\rm e}^{-\frac{(q_1-x_1)^2}{2\Delta_1^2}}{\rm e}^{-\frac{(q'_2-q_1+x_2)^2}{2\Delta_2^2}}+{\rm e}^{-\frac{(q_1-x_2)^2}{2\Delta_1^2}} {\rm e}^{-\frac{(q'_2-q_1+x_1)^2}{2\Delta_2^2}}  \right)\ket{q_2}\bra{q'_2}\ ,
\end{split}
\end{equation}
so that, after evaluating the integration in $d q_1$, the probability $P(y)=\mel{y}{\Tr_1\left(\ket{\tilde{\psi}_{in}}\bra{\tilde{\psi}_{in}}\right)}{y}$, becomes \eqref{eq:PyXOR}
\begin{equation}
P(y)=\frac{1}{C \sqrt{\pi (\Delta_1^2+\Delta_2^2)} }\left( {\rm e}^{-\frac{(y-x_1+x_2)^2}{\Delta_1^2+\Delta_2^2}} + {\rm e}^{-\frac{(y-x_2+x_1)^2}{\Delta_1^2+\Delta_2^2}} + 2{\rm e}^{-\frac{(x_1-x_2)^2(\Delta_1^2+\Delta_2^2)}{4\Delta_1^2\Delta_2^2}}{\rm e}^{-\frac{y^2}{\Delta_1^2+\Delta_2^2}}\right)\ .
\end{equation}
\vskip 1cm

\noindent
\textbf{Acknowledgements}\quad The author F.B. acknowledges that his research has been conducted within the framework of the Trieste Institute for Theoretical Quantum Technologies.

\bibliographystyle{plain}

\end{document}